\documentclass
[preprint,prl,notitlepage,twocolumn,11pt,showpacs,tightenlines,secnumarabic,nofootinbib]{revtex4}%
\usepackage{amssymb}
\usepackage{amsfonts}
\usepackage{amsmath}
\usepackage{graphicx}%
\providecommand{\U}[1]{\protect\rule{.1in}{.1in}}

\begin{document}
\preprint{ }
\title[Born-Jordan quantization]{The Angular Momentum Dilemma and Born--Jordan Quantization}
\author{Maurice A. de Gosson}
\affiliation{University of Vienna, Institute of Mathematics, NuHAG}
\keywords{Born--Jordan quantization, entanglement, phase space picture}
\pacs{02.30.Nw, 03.65.Ca, 03.67.Bg, 03.65.Ta}

\begin{abstract}
We have shown in previous work that the rigorous equivalence of the
Schr\"{o}dinger and Heisenberg pictures requires that one uses Born--Jordan
quantization in place of Weyl quantization. It also turns out that the
so-called angular momentum dilemma disappears if one uses Born--Jordan
quantization. These two facts strongly suggest that the latter is the\ only
true quantization procedure, and this leads to a redefinition of phase space
quantum mechanics, where the usual Wigner distribution has to be replaced with
a new distribution.

\end{abstract}
\volumeyear{year}
\volumenumber{number}
\issuenumber{number}
\eid{identifier}
\date[Date text]{date}
\received[Received text]{date}

\revised[Revised text]{date}

\accepted[Accepted text]{date}

\published[Published text]{date}

\startpage{101}
\endpage{105}
\maketitle

\section{Introduction}

To address quantization problems in these \textquotedblleft Times of
Entanglement\textquotedblright\ is not very fashionable: everything seems to
have been said about this old topic, and there is more or less a consensus
about the best way to quantize a physical system: it should be done using the
Weyl transformation. The latter, in addition to being relatively simple,
enjoys several nice properties, one of the most important being its
\textquotedblleft symplectic covariance\textquotedblright, reflecting, at the
quantum level, the canonical covariance of Hamiltonian dynamics. Things are,
however, not that simple. If one insists in using Weyl quantization, one gets
inconsistency, because the Schr\"{o}dinger and Heisenberg pictures are then
not equivalent. Dirac already notes in the Abstract to his paper \cite{PAM}
that \textquotedblleft...\textit{the Heisenberg picture is a good picture, the
Schr\"{o}dinger picture is a bad picture, and the two pictures are not
equivalent}...\textquotedblright. This observation has also been confirmed by
Kauffmann's \cite{Kauffmann} interesting discussion of the non-physicality of
Weyl quantization.

This non-physicality is made strikingly explicit on an annoying contradiction
known as the \textquotedblleft angular momentum dilemma\textquotedblright: the
Weyl quantization of the squared classical angular-momentum is not the squared
quantum angular momentum operator, but it contains an additional term
$\frac{3}{2}\hbar^{2}$. This extra term is actually physically significant,
since it accounts for the nonvanishing angular momentum of the ground-state
Bohr orbit in the hydrogen atom. This contradiction has been noted by several
authors\footnote{It is also mentioned in Wikipedia's article \cite{wiki} on
geometric quantization.}, to begin with Linus Pauling's in his \textit{General
Chemistry }\cite{Pauling}; it is also taken up by Shewell \cite{Shewell}, and
discussed by Dahl and Springborg \cite{dasp} and by Dahl and Schleich
\cite{das}.

It turns out that we have shown in a recent work \cite{FP} that the
Schr\"{o}dinger and Heisenberg pictures cannot be equivalent unless we use a
quantization rule proposed by Born and Jordan's \cite{bj,bjh}, and which
precedes Weyl's rule \cite{Weyl} by almost two years. This suggests that Weyl
quantization should be replaced with Born--Jordan (BJ) quantization in the
Schr\"{o}dinger picture. Now, at first sight, this change of quantization
rules does not lead to any earthshaking consequences for the Schr\"{o}dinger
picture, especially since one can prove \cite{TRANSAM} that BJ and Weyl
quantizations coincide for all Hamiltonian functions of the type
\textquotedblleft kinetic energy + potential\textquotedblright\ or, more
generally, for Hamiltonian functions of the type%
\[
H(x,p)=\sum_{j=1}^{n}\frac{1}{2m_{j}}(p_{j}-A_{j}(x))^{2}+V(x)
\]
even when the potentials $A=(A_{1},..,A_{n})$ and $V$ are irregular (we are
using generalized coordinates $x=(x_{1},...,x_{n})$, $p=(p_{1},...,p_{n})$).
One can therefore wonder whether it is really necessary to write yet another
paper on quantization rules, just to deal with a quasi-philosophical problem
(the equivalence of two pictures of quantum mechanics) and one little anomaly
(the angular momentum dilemma, we will discuss below). The quantum world is
however more subtle than that. The problem is that if we stick to Weyl
quantization for general systems, another inconsistency appears, which has
far-reaching consequences. It is due to the fact the commonly used phase space
picture of quantum mechanics, where the Wigner distribution plays a central
role, is intimately related to Weyl quantization. In short, if we change the
quantization rules, we also have to change the phase space picture, thus
leading not only to a redefinition of the Wigner distribution, but also to
substantial changes in related phase space objects, such as, for instance the
Moyal product of two observables, which is at the heart of deformation
quantization; these aspects are discussed in detail in \cite{Springer}.

\section{BJ versus Weyl: the case of monomials}

Born and Jordan (BJ) proved in \cite{bjh} that the only way to quantize
polynomials in a way consistent with Heisenberg's ideas was to use the rule%
\begin{equation}
p_{j}^{s}x_{j}^{r}\overset{\mathrm{BJ}}{\longrightarrow}\frac{1}{s+1}%
\sum_{\ell=0}^{s}\widehat{p}_{j}^{s-\ell}\widehat{x}_{j}^{r}\widehat{p}%
_{j}^{\ell}; \label{bj1}%
\end{equation}
or, equivalently,
\begin{equation}
p_{j}^{s}x_{j}^{r}\overset{\mathrm{BJ}}{\longrightarrow}\frac{1}{r+1}%
\sum_{j=0}^{r}\widehat{x}_{j}^{r-j}\widehat{p}_{j}^{s}\widehat{x}_{j}^{j}.
\label{bj2}%
\end{equation}
The BJ quantization is thus the equally weighted average of all the possible
operator orderings. Weyl \cite{Weyl} proposed, independently, some time later
a very general rule: elaborating on the Fourier inversion formula, he proposed
that the quantization $\widehat{A}$ of a classical observable $a(x,p)$ should
be given by
\[
\widehat{A}=\left(  \tfrac{1}{2\pi\hbar}\right)  ^{n}\int Fa(x,p)e^{\frac
{i}{\hbar}(x\widehat{x}+p\widehat{p})}d^{n}xd^{n}p
\]
where $Fa(x,p)$ is the Fourier transform of $a(x,p)$; applying this rule to
monomials yields (McCoy \cite{mccoy})
\begin{equation}
p_{j}^{s}x_{j}^{r}\overset{\mathrm{Weyl}}{\longrightarrow}\frac{1}{2^{s}}%
\sum_{\ell=0}^{s}%
\begin{pmatrix}
s\\
\ell
\end{pmatrix}
\widehat{p}_{j}^{s-\ell}\widehat{x}_{j}^{r}\widehat{p}_{j}^{\ell}. \label{w2}%
\end{equation}
It turns out that both the BJ and Weyl rules coincide as long as $s+r\leq2$,
but they are different as soon as $s\geq2$ and $r\geq2$. For instance, to $xp$
corresponds $\frac{1}{2}(\widehat{x}\widehat{p}+\widehat{p}\widehat{x}$) in
both cases, but
\begin{align*}
&  x^{2}p^{2}\overset{\mathrm{BJ}}{\longrightarrow}\frac{1}{3}(\widehat{x}%
^{2}\widehat{p}^{2}+\widehat{x}\widehat{p}^{2}\widehat{x}+\widehat{p}%
^{2}\widehat{x}^{2})\\
&  x^{2}p^{2}\overset{\mathrm{Weyl}}{\longrightarrow}\frac{1}{4}(\widehat
{x}^{2}\widehat{p}^{2}+2\widehat{x}\widehat{p}^{2}\widehat{x}+\widehat{p}%
^{2}\widehat{x}^{2})
\end{align*}
and both expressions differ by the quantity $\frac{1}{2}\hbar^{2}$ as is
easily checked by using several times the commutation rule $[\widehat
{x},\widehat{p}]=i\hbar$. We now make the following essential remark: let
$\tau$ be a real number, and consider the somewhat exotic quantization rule%
\begin{equation}
p_{j}^{s}x_{j}^{r}\overset{\tau}{\longrightarrow}\sum_{\ell=0}^{s}%
\begin{pmatrix}
s\\
\ell
\end{pmatrix}
(1-\tau)^{\ell}\tau^{s-\ell}\widehat{p}_{j}^{s-\ell}\widehat{x}_{j}%
^{r}\widehat{p}_{j}^{\ell}. \label{tau1}%
\end{equation}
Clearly, the choice $\tau=\frac{1}{2}$ immediately yields Weyl's rule
(\ref{w2}); what is less obvious is that if we integrate the right-hand side
of (\ref{tau1}), then we get at once the BJ rule (\ref{bj2}). This remark
allows us to define BJ quantization for arbitrary observables.

\section{Generalized BJ Quantization}

Our definition is inspired by earlier work in signal theory by Boggiato and
his coworkers \cite{bogetal,bogetalbis}; for the necessary background in Weyl
correspondence we refer to Littlejohn's seminal paper \cite{Littlejohn}, whose
notation we use here (also see \cite{Birkbis}). Let $a=a(x,p)$ be an
observable; writing $z=(x,p)$ the Weyl operator $\widehat{A}%
=\operatorname*{Op}_{\mathrm{W}}(a)$ is defined by%
\begin{equation}
\widehat{A}\psi=\left(  \tfrac{1}{2\pi\hbar}\right)  ^{n}\int a_{\sigma
}(z)\widehat{T}(z)\psi d^{2n}z \label{Weyl}%
\end{equation}
where $d^{2n}z=dx_{1}\cdot\cdot\cdot dx_{n}dp_{1}\cdot\cdot\cdot dp_{n}$ and
\begin{equation}
a_{\sigma}(z)=\left(  \tfrac{1}{2\pi\hbar}\right)  ^{n}\int e^{-\frac{i}%
{\hbar}\sigma(z,z^{\prime})}a(z)d^{2n}z \label{asigma}%
\end{equation}
is the symplectic Fourier transform of\ $a$. Here $\widehat{T}(z)=e^{-i\sigma
(\widehat{z},z)/\hbar}$ is the Heisenberg--Weyl operator; $\sigma$ is the
standard symplectic form defined by $\sigma(z,z^{\prime})=px^{\prime
}-p^{\prime}x$ if $z=(x,p)$, $z^{\prime}=(x^{\prime},p^{\prime})$. The natural
generalization of the $\tau$-rule (\ref{tau1}) is obtained
\cite{TRANSAM,golu1} by replacing $\widehat{T}(z)$ with $=e^{-i\sigma_{\tau
}(\widehat{z},z)/\hbar}$ where%
\[
\widehat{T}_{\tau}(z)=\exp\left(  \frac{i}{2\hbar}(2\tau-1)px\right)
\widehat{T}(z);
\]
integrating for $0\leq\tau\leq1$ one gets the Born--Jordan operator associated
with the observable $a$:%
\begin{equation}
A_{\mathrm{BJ}}\psi=\left(  \tfrac{1}{2\pi\hbar}\right)  ^{n}\int a_{\sigma
}(z)\Theta(z)\widehat{T}(z)\psi d^{2n}z \label{BJ}%
\end{equation}
where%
\begin{equation}
\Theta(z)=\int_{0}^{1}e^{\frac{i}{2\hbar}(2\tau-1)px}d\tau=\frac
{\sin(px/2\hbar)}{px/2\hbar} \label{sinc}%
\end{equation}
with $px=p_{1}x_{1}+\cdot\cdot\cdot+p_{n}x_{n}$. We thus have
\begin{equation}
(a_{\mathrm{W}})_{\sigma}(x,p)=a_{\sigma}(x,p)\frac{\sin(px/2\hbar)}%
{px/2\hbar}. \label{awbis}%
\end{equation}
Taking the symplectic Fourier transform of $a_{\sigma}(z)\Theta(z)$, this
means that the Weyl transform of $A_{\mathrm{BJ}}$ is the phase space
function
\begin{equation}
a_{\mathrm{W}}=\left(  \tfrac{1}{2\pi\hbar}\right)  ^{n}a\ast\Theta_{\sigma}.
\label{aw}%
\end{equation}
The appearance of the function $\Theta$ in the formulas above is interesting;
we have
\[
\Theta(z)=\operatorname{sinc}\left(  \frac{px}{2\hbar}\right)
\]
where $\operatorname{sinc}$ is Whittaker's \textit{sinus cardinalis}
function\ familiar from Fraunhofer diffraction \cite{bw}\footnote{I thank
Basil Hiley for having drawn my attention to this fact.}.

We now make an important remark: suppose that we split the phase space point
$(x,p)$ into two sets of independent coordinates $z^{\prime}=(x^{\prime
},p^{\prime})$ and $z^{\prime\prime}=(x^{\prime\prime},p^{\prime\prime})$. Let
$b(z^{\prime})$ be an observable in the first set, and $c(z^{\prime\prime})$
an observable in the second set, and define $a=b\otimes c$; it is an
observable depending on the total set variables $(x,p)=(x^{\prime}%
,x^{\prime\prime},p^{\prime},p^{\prime\prime})$. We obviously have $a_{\sigma
}=b_{\sigma}\otimes c_{\sigma}$ and $\widehat{T}(z)=\widehat{T}(z^{\prime
})\otimes\widehat{T}(z^{\prime\prime})$ (because the symplectic form $\sigma$
splits in the sum $\sigma^{\prime}\oplus$ $\sigma^{\prime\prime}$ of the two
standard symplectic forms $\sigma^{\prime}$ and $\sigma^{\prime\prime}$
defined on, respectively, $z^{\prime}$ and $z^{\prime\prime}$ phase spaces).
It follows from formula (\ref{Weyl}) that we have $\widehat{A}=\widehat
{B}\otimes\widehat{C}$; \textit{i.e.} Weyl quantization preserves the
separation of two observables. This property is generically not true for
Born--Jordan quantization: because of the presence in formula (\ref{BJ}) of
the function $\Theta(z)$ we cannot write the integrand as a tensor product,
and hence we have in general%
\begin{equation}
A_{\mathrm{BJ}}\neq B_{\mathrm{BJ}}\otimes C_{\mathrm{BJ}} \label{abjc}%
\end{equation}
In this sense, Born--Jordan quantization \textquotedblleft
entangles\textquotedblright\ quantum observables.

\section{The angular momentum dilemma}

Dahl and Springborg's \cite{dasp} argument we alluded to in the introduction
boils down to the following observation for the electron in the hydrogen atom
in its $1s$ state. Let
\begin{equation}
\widehat{\ell}=(\widehat{x}_{2}\widehat{p}_{3}-\widehat{x}_{3}\widehat{p}%
_{2})\mathbf{i}+(\widehat{x}_{3}\widehat{p}_{1}-\widehat{x}_{1}\widehat{p}%
_{3})\mathbf{j}+(\widehat{x}_{1}\widehat{p}_{2}-\widehat{x}_{2}\widehat{p}%
_{1})\mathbf{k} \label{amp}%
\end{equation}
be the angular momentum operator and
\begin{equation}
\widehat{\ell}^{2}=(\widehat{x}_{2}\widehat{p}_{3}-\widehat{x}_{3}\widehat
{p}_{2})^{2}+(\widehat{x}_{3}\widehat{p}_{1}-\widehat{x}_{1}\widehat{p}%
_{3})^{2}+(\widehat{x}_{1}\widehat{p}_{2}-\widehat{x}_{2}\widehat{p}_{1})^{2}
\label{ampsqr}%
\end{equation}
its square. According to the Bohr model, the square of the classical angular
momentum%
\begin{equation}
\ell=(x_{2}p_{3}-x_{3}p_{2},x_{3}p_{1}-x_{1}p_{3},x_{1}p_{2}-x_{2}p_{1})
\label{angular}%
\end{equation}
should have the value $\hbar^{2}$, while it is zero in the Schr\"{o}dinger
picture. Thus, Dahl and Springborg contend, the \textquotedblleft
dequantization\textquotedblright\ of $\widehat{\ell}^{2}$ should yield the
Bohr value\footnote{Of course, their argument is heuristic, because there is
\textit{in general} no relation between the eigenvalues of a quantum operator
and the values of the corresponding classical observable.}. Now,
\textquotedblleft dequantizing\textquotedblright\ $\widehat{\ell}^{2}$ using
the Weyl transformation leads to the function $\ell^{2}+\tfrac{3}{2}\hbar^{2}$
(as already remarked by Shewell \cite{Shewell}, formula (4.10)), which gives
the \textquotedblleft wrong\textquotedblright\ value $\tfrac{3}{2}\hbar^{2}$
for the Bohr angular momentum. However, if we view $\widehat{\ell}^{2}$ as the
Born--Jordan quantization of $\ell^{2}$, then we recover the Bohr value
$\hbar^{2}$. Let us show this in some detail. It suffices of course to study
one of the three terms appearing in the square of the vector (\ref{angular}),
say
\begin{equation}
\ell_{3}^{2}=x_{1}^{2}p_{2}^{2}+x_{2}^{2}p_{1}^{2}-2x_{1}p_{1}x_{2}p_{2}.
\label{l2}%
\end{equation}
The two first terms in (\ref{l2}) immediately yield the operators $\widehat
{x}_{1}^{2}\widehat{p}_{2}^{2}$ and $\widehat{x}_{2}^{2}\widehat{p}_{1}^{2}$
(as they would in any realistic quantization scheme), so let us focus on the
third term $a_{12}(z)=2x_{1}p_{1}x_{2}p_{2}$ (we are writing here
$z=(x_{1},x_{2},p_{1},p_{2})$). Using the standard formula giving the Fourier
transform of a monomial we get%
\begin{equation}
a_{12,\sigma}(z)=2\hbar^{4}(2\pi\hbar)^{2}\delta^{\prime}(z) \label{asigdelta}%
\end{equation}
where we are using the notation
\begin{align*}
\delta(z)  &  \equiv\delta(x_{1})\otimes\delta(x_{2})\otimes\delta
(p_{1})\otimes\delta(p_{2}),\\
\delta^{\prime}(z)  &  \equiv\delta^{\prime}(x_{1})\otimes\delta^{\prime
}(x_{2})\otimes\delta^{\prime}(p_{1})\otimes\delta^{\prime}(p_{2}).
\end{align*}
Expanding the function $\sin(px/2\hbar)$ in a Taylor series, we get%
\[
\Theta(z)=1+\sum_{k=1}^{\infty}\frac{(-1)^{k}}{(2k+1)!}\left(  \frac
{px}{2\hbar}\right)  ^{2k}%
\]
and hence, observing that $(px)^{2k}\delta^{\prime}(z)=0$ for $k>1$,
\begin{equation}
a_{12,\sigma}(z)\Theta(z)=a_{12,\sigma}(z)\left(  1-\frac{(px)^{2}}%
{24\hbar^{2}}\right)  . \label{F}%
\end{equation}
Comparing the expressions (\ref{Weyl}) and (\ref{BJ}), defining respectively
the Weyl and Born--Jordan quantizations of $a$, it follows that the
difference
\[
\Delta(a_{12})=\operatorname*{Op}\nolimits_{\mathrm{BJ}}(a_{12}%
)-\operatorname*{Op}\nolimits_{\mathrm{W}}(a_{12})
\]
is given by%
\begin{align*}
\Delta(a_{12})\psi &  =\frac{1}{24\hbar^{2}}\left(  \frac{1}{2\pi\hbar
}\right)  ^{2}\int a_{12,\sigma}(z)(px)^{2}\widehat{T}(z)\psi d^{4}z\\
&  =\frac{\hbar^{2}}{12}\int\delta^{\prime}(z)(px)^{2}\widehat{T}(z)\psi
d^{4}z.
\end{align*}
Using the elementary properties of the Dirac function we have
\begin{equation}
\delta^{\prime}(z)(px)^{2}=2\delta(z) \label{diz}%
\end{equation}
and hence%
\[
\Delta(a_{12})\psi=\frac{\hbar^{2}}{6}\int\delta(z)\widehat{T}(z)\psi
d^{4}z=\frac{\hbar^{2}}{6}\psi
\]
the second equality because
\[
\delta(z)\widehat{T}(z)=\delta(z)e^{-\frac{i}{\hbar}\sigma(\widehat{z}%
,z)}=\delta(z).
\]
A similar calculation for the quantities $\Delta(a_{23})$\ and $\Delta
(a_{13})$ corresponding to terms $\ell_{1}^{2}$ and $\ell_{2}^{2}$ leads to
the formula%
\begin{equation}
\operatorname*{Op}\nolimits_{\mathrm{BJ}}(\ell^{2})-\operatorname*{Op}%
\nolimits_{\mathrm{W}}(\ell^{2})=\tfrac{1}{2}\hbar^{2}, \label{opop2}%
\end{equation}
hence, taking (\ref{ampsqr}) into account:
\begin{equation}
\operatorname*{Op}\nolimits_{\mathrm{BJ}}(\ell^{2})=\widehat{\ell}^{2}%
+\hbar^{2} \label{19}%
\end{equation}
which is the expected result. We note that Dahl and Springborg \cite{dasp} get
the same operator $\widehat{\ell}^{2}+\hbar^{2}$ by averaging the Weyl
operator $\operatorname*{Op}\nolimits_{\mathrm{W}}(\ell^{2})$ over what they
call a \textquotedblleft classical subspace\textquotedblright; funnily enough
the $\operatorname{sinc}$ function also appears at some moment in their
calculations (formula (40)). It would be interesting to see whether this is a
mere coincidence, or if a hidden relation with BJ quantization already is
involved in these calculations.

\section{The BJ-Wigner Transform}

As we mentioned in the introduction, the phase space picture very much depends
on the used quantization. In the Wigner formalism, if $A_{\mathrm{W}%
}=\operatorname*{Op}_{\mathrm{W}}(a)$,%
\begin{equation}
\langle\psi|A_{\mathrm{W}}|\psi\rangle=\int a(z)W\psi(z)d^{2n}z\label{aw2}%
\end{equation}
where $\psi$ is normalized, and
\[
W\psi(z)=\left(  \tfrac{1}{2\pi\hbar}\right)  ^{n}\int e^{-\frac{i}{\hbar}%
py}\psi(x+\tfrac{1}{2}y)\psi^{\ast}(x-\tfrac{1}{2}y)d^{n}y
\]
is the usual Wigner quasi distribution \cite{Birk,Birkbis,Littlejohn}. As we
have shown in \cite{TRANSAM}, if we replace Weyl quantization of the classical
observable $a$ with its Born--Jordan quantization $A_{\mathrm{BJ}%
}=\operatorname*{Op}_{\mathrm{BJ}}(a)$, then formula (\ref{aw2}) becomes
\begin{equation}
\langle\psi|A_{\mathrm{BJ}}|\psi\rangle=\int a(z)W_{\mathrm{BJ}}\psi
(z)d^{2n}z\label{abj2}%
\end{equation}
where $W_{\mathrm{BJ}}\psi$ is given by the convolution formula%
\begin{equation}
W_{\mathrm{BJ}}\psi(z)=\left(  \tfrac{1}{2\pi\hbar}\right)  ^{n}W\psi
\ast\Theta_{\sigma}\label{wbj}%
\end{equation}
(this can easily be proven using formula (\ref{aw})). Let us compare the
expressions $\langle\psi|A_{\mathrm{W}}|\psi\rangle$ and $\langle
\psi|A_{\mathrm{BJ}}|\psi\rangle$ where $A_{\mathrm{W}}=\operatorname*{Op}%
_{\mathrm{W}}(a)$ and $A_{\mathrm{BJ}}=\operatorname*{Op}_{\mathrm{BJ}}(a)$.
In view of Parseval's theorem we can rewrite formulas (\ref{aw2}) and
(\ref{abj2}) as%
\begin{align}
\langle\psi|A_{\mathrm{W}}|\psi\rangle &  =\int a_{\sigma}(z)F_{\sigma}%
W\psi(z)d^{2n}z\label{awpsi}\\
\langle\psi|A_{\mathrm{BJ}}|\psi\rangle &  =\int a_{\sigma}(z)F_{\sigma
}W_{\mathrm{BJ}}\psi(z)d^{2n}z.\label{abjpsi}%
\end{align}
Since $F_{\sigma}W_{\mathrm{BJ}}\psi(z)=F_{\sigma}W\psi(z)\Theta(z)$ we have%
\begin{multline*}
\langle\psi|A_{\mathrm{W}}|\psi\rangle-\langle\psi|A_{\mathrm{BJ}}|\psi
\rangle=\\
\int a_{\sigma}(z)F_{\sigma}W\psi(z)(1-\Theta(z))d^{2n}z.
\end{multline*}
Let us apply this formula to the square $\ell^{2}$ of the angular momentum. As
above, we only have to care about the cross term (\ref{asigdelta}); in view of
formula (\ref{F}) above we have%
\[
a_{\sigma}(z)(1-\Theta(z))=\frac{\hbar^{2}}{6}(2\pi\hbar)^{2}\delta(z)
\]
and hence%
\begin{multline*}
\langle\psi|A_{\mathrm{W}}|\psi\rangle-\langle\psi|A_{\mathrm{BJ}}|\psi
\rangle=\\
\frac{\hbar^{2}}{6}(2\pi\hbar)^{2}\int\delta(z)F_{\sigma}W\psi(0)dz.
\end{multline*}
Observing that
\[
F_{\sigma}W\psi(0)=\left(  \tfrac{1}{2\pi\hbar}\right)  ^{2}\int
W\psi(z)dz=\left(  \tfrac{1}{2\pi\hbar}\right)  ^{2}%
\]
we finally get%
\[
\langle\psi|A_{\mathrm{BJ}}|\psi\rangle-\langle\psi|A_{\mathrm{W}}|\psi
\rangle=-\tfrac{1}{6}\hbar^{2}.
\]
it follows, taking the two other cross-components of\ $\ell^{2}$ into account,
that we have%
\[
\langle\psi|\ell_{\mathrm{BJ}}^{2}|\psi\rangle-\langle\psi|\ell_{\mathrm{W}%
}^{2}|\psi\rangle=-\tfrac{1}{2}\hbar^{2}%
\]
and hence
\begin{equation}
\langle\psi|\ell_{\mathrm{BJ}}^{2}|\psi\rangle=\hbar^{2}\label{accord}%
\end{equation}
as predicted by Bohr's theory.

\section{Discussion}

We have tried to make it clear that to avoid inconsistencies one has to use BJ
quantization instead of the Weyl correspondence in the Schr\"{o}dinger picture
of quantum mechanics. With hindsight, it somewhat ironic that the
\textquotedblleft true\textquotedblright\ quantization should be the one which
was historically the first to be proposed. There are, however, unexpected
difficulties that appear; the mathematics of BJ quantization is not fully
understood. For instance, the generalized Born--Jordan rule (\ref{BJ}) does
not implement a true \textquotedblleft correspondence\textquotedblright, as
the Weyl rule does. In fact, it results from a deep mathematical theorem,
Schwartz's kernel theorem \cite{Birkbis}, that every quantum observable $A$
which is sufficiently smooth can be viewed as the Weyl transform of some
classical observable $a$, and conversely. However, this is not true of the BJ
quantization scheme: it is not true that to every quantum observable (or
\textquotedblleft operator\textquotedblright) one can associate a classical
observable. In fact, rewriting formula (\ref{aw}) as
\begin{equation}
(a_{\mathrm{W}})_{\sigma}(x,p)=a_{\sigma}(x,p)\frac{\sin(px/2\hbar)}%
{px/2\hbar}\label{awter}%
\end{equation}
we see that we cannot, in general, calculate $a_{\sigma}(x,p)$ (and hence
$a(x,p)$) if we know the Weyl transform $a_{\mathrm{W}}$ of $A$, and this
because the function $\Theta(x,p)=\sin(px/2\hbar)/(px/2\hbar)$ has infinitely
many zeroes: $\Theta(x,p)=0$ for all phase space points $(x,p)$ such that
$p_{1}x_{1}+\cdot\cdot\cdot+p_{n}x_{n}=0$. We are thus confronted with a
difficult division problem; see \cite{cogoni}. It is also important to note
that we loose uniqueness of quantization when we use the Born--Jordan
\textquotedblleft correspondence\textquotedblright: if $(a_{\mathrm{W}%
})_{\sigma}(x,p)=0$ there are infinitely many Weyl operators who verify
(\ref{awter}). These issues, which might lead to interesting developments in
quantum mechanics, are discussed in detail in our book \cite{Springer}. The
BJ-Wigner transform and its relation with what we call \textquotedblleft
Born--Jordan quantization\textquotedblright\ has been discovered independently
by Boggiatto and his coworkers \cite{bogetal,bogetalbis} who were working on
certain questions in signal theory and time-frequency analysis; they show --
among other things -- that the spectrograms obtained by replacing the standard
Wigner distribution by its modified version $W_{\mathrm{BJ}}\psi$ are much
more accurate. The properties of $W_{\mathrm{BJ}}\psi$ are very similar to
those of $W\psi$; it is always a real function, and it has the
\textquotedblleft right\textquotedblright\ marginals and can thus be treated
as a quasi-distribution, exactly as the traditional Wigner distribution does.

\end{document}